\documentclass{article}
\usepackage{spconf,amsmath,graphicx,hyperref}
\usepackage{multirow}
\usepackage{makecell}
\usepackage{booktabs}
\usepackage{xcolor}
\usepackage{threeparttable}
\usepackage[inline]{enumitem}
\usepackage{cleveref}

\title{ACAVCaps: Enabling Large-Scale Training for Fine-Grained and Diverse Audio Understanding}
\name{
\begin{tabular}[t]{@{}c@{}}
Yadong Niu\textsuperscript{1}, Tianzi Wang\textsuperscript{1,2}, Heinrich Dinkel\textsuperscript{1}, Xingwei Sun\textsuperscript{1}, Jiahao Zhou\textsuperscript{1}, \\
Gang Li\textsuperscript{1}, Jizhong Liu\textsuperscript{1}, Junbo Zhang\textsuperscript{1}, Jian Luan\textsuperscript{1}
\end{tabular}
}
\address{
\textsuperscript{1} MiLM Plus, Xiaomi Inc, Beijing, China\\
\textsuperscript{2} The Chinese University of Hong Kong, Hong Kong, China\\
}
\begin{document}

\maketitle
%
\abstract
General audio understanding is a fundamental goal for large audio-language models, with audio captioning serving as a cornerstone task for their development. However, progress in this domain is hindered by existing datasets, which lack the scale and descriptive granularity required to train truly versatile models. To address this gap, we introduce ACAVCaps, a new large-scale, fine-grained, and multi-faceted audio captioning dataset. 
Derived from the ACAV100M collection, ACAVCaps is constructed using a multi-expert pipeline that analyzes audio from diverse perspectives—including speech, music, and acoustic properties—which are then synthesized into rich, detailed descriptions by a large language model. 
Experimental results demonstrate that models pre-trained on ACAVCaps exhibit substantially stronger generalization capabilities on various downstream tasks compared to those trained on other leading captioning datasets. 
The dataset is avaiable at https://github.com/xiaomi-research/acavcaps.
\endabstract
\vspace{-1mm}
\section{Introduction}
\label{sec:intro}

The ability to comprehend complex acoustic environments is a fundamental goal in the pursuit of general artificial intelligence. 
Large Audio-Language Models (LALMs) have recently emerged as a promising paradigm for this task, aiming to build models with a deep and versatile understanding of sound. 
A cornerstone for developing such models is the task of audio captioning—generating rich, human-readable descriptions of acoustic content, which serves as a powerful bridge between the audio and text modalities. 

Despite their rapid progress, the performance of current LALMs on diverse, real-world audio understanding tasks remains constrained. 
We argue that this limitation stems not from the models themselves, but from the data they are trained on.
Existing audio captioning datasets suffer from several critical issues: 
\begin{enumerate*}[label=(\Roman*)]
\item Scarcity of high-Fidelity data at scale. Creating large-scale datasets with accurate and detailed descriptions is inherently challenging. Manual annotation is costly and difficult to scale;
\item Homogeneous Sources and Rigid Annotations. Most large-scale datasets are confined to limited domains (e.g., AudioSet) and characterized by a single, stylistic pattern that lacks the linguistic variety found in the real world;
\item Lack of Descriptive Granularity. Captions are often too generic (e.g., ``a man is speaking'') and fail to provide the discriminative acoustic features needed to distinguish between nuanced auditory events. 
\end{enumerate*} 
Effective audio-text alignment is crucial for LALMs, which requires a training dataset that provides a fine-grained mapping from audio to descriptive text. 
High-quality, detailed captions enable this, allowing models to learn generalizable representations for diverse tasks.

We introduce ACAVCaps, a large-scale, fine-grained audio captioning dataset derived from ACAV100M~\cite{lee2021acav100m}. 
To create detailed captions, we analyze audio with specialized expert models and synthesize their outputs using a Chain-of-Thought (CoT) enhanced large language model (LLM). 
This pipeline produces richly detailed descriptions to train more capable audio models.

\begin{figure*}[h]
    \centering
    \includegraphics[width=\linewidth]{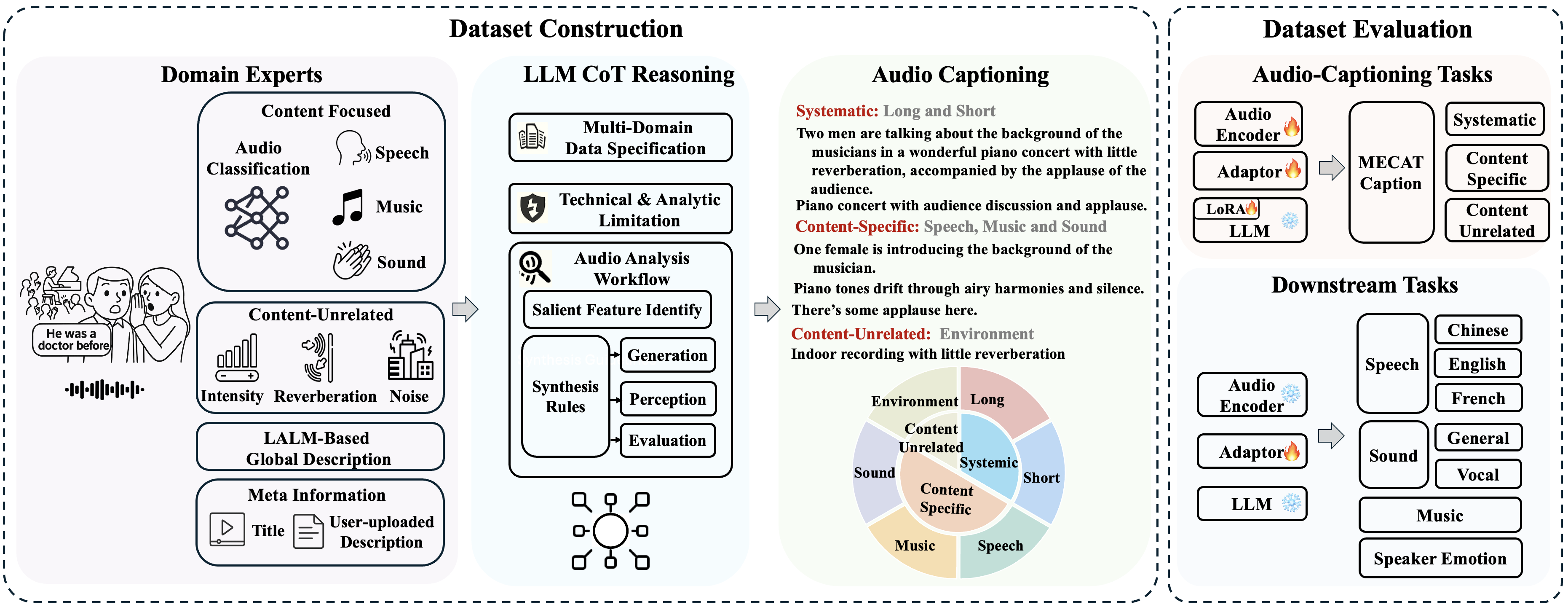}
    \caption{Data construction and evaluation frameowrk.}
    \label{fig:labeling-main}
    \vspace{-3mm}
\end{figure*}

\begin{table*}[t]
\centering
\caption{Comparison of ACAVCaps with existed caption datasets. The Unique Tokens column reports the total number of unique tokens within each dataset, as counted by the Qwen3 tokenizer. ~$\dagger$ MP-LLM: Multiple Experts Models and LLM; $\ddagger$ Multi-Domain: This includes speech, music and sound-events ($\diamond$ denotes that domain were not elaborated in detail); $\S$ Extended Multi-Domain: This includes speech, music, sound-events, combinations thereof, and silence.}
\label{tab:caption_comparison}
\begin{tabular}{lllllll}
\toprule
Labeling & Dataset & Duration (h) & Samples & Unique Tokens & Domain & Source \\
\midrule
\multirow{3}{*}{\makecell[l]{Manual}} & 
AudioCaps~\cite{kim2019audiocaps} & 135 & 50k & 5.5k & Multi-Domain\textsuperscript{$\ddagger$,$\diamond$} & AudioSet \\
& Clotho~\cite{drossos2020clotho} & 24 & 3.8k & 5.5k & Multi-Domain\textsuperscript{$\ddagger$,$\diamond$} & FreeSound \\
& SongDescriber~\cite{manco2023song} & 12 & 0.4k & 2.4k & Music & MTG-Jamendo \\
\cmidrule(lr){1-7} 
\multirow{6}{*}{\makecell[l]{LLM}} & 
MusicCaps~\cite{agostinelli2023music_caps} & 7 & 4.6k & 4.6k & Music & AudioSet \\
& LPMusicCaps~\cite{doh2023lpmusiccaps} & 127 & 21.6k & 5.1k & Music & Audioset, MSD \\
& WavCaps~\cite{mei2024wavcaps} & 1.8k & 0.4M & 23.1k & Multi-Domain\textsuperscript{$\ddagger$,$\diamond$}  & AudioSet \\
& Auto-ACD~\cite{sun2024auto} & 5.2k & 1.9M & 20.3k & Multi-Domain\textsuperscript{$\ddagger$,$\diamond$} & AudioSet \\
& Sound-VeCaps~\cite{yuan2025sound} & 4.5k & 1.6M & 42.7k & Multi-Domain\textsuperscript{$\ddagger$,$\diamond$} & AudioSet \\
& AudioSetCaps~\cite{bai2025audiosetcaps} & 5.6k & 2.0M & 20.9k & Multi-Domain\textsuperscript{$\ddagger$,$\diamond$} & AudioSet \\
\cmidrule(lr){1-7} 
\multirow{1}{*}{\makecell[l]{MP-LLM\textsuperscript{$\dagger$}}} & 
\textbf{ACAVCaps (Ours)} & \textbf{13.0k} & \textbf{4.7M} & \textbf{76.7k} & Extended~Multi-Domain\textsuperscript{$\S$} & ACAV100M \\
\bottomrule
\end{tabular}%
\vspace{-3mm}
\end{table*}

\vspace{-3mm}
\section{Related works} 
The advancement of general audio understanding is intrinsically linked to the quality and diversity of training datasets. 
The field has evolved from smaller, manually annotated corpora to large-scale, automatically generated ones, yet significant challenges related to data scale, descriptive granularity, and source limitations persist across all major audio domains.

In the domain of general sound events, early foundational datasets like AudioCaps~\cite{kim2019audiocaps} and Clotho~\cite{drossos2020clotho} were created through intensive manual annotation. 
While providing high-quality human descriptions, their data scale is inherently limited (typically a few thousand samples), and their captions often lack fine-grained detail, focusing on generic, event-level descriptions. 
To address the scale issue, a new wave of datasets was created using automated pipelines. 
These include WavCaps~\cite{mei2024wavcaps}, AudioSetCaps~\cite{bai2025audiosetcaps}, and Auto-ACD~\cite{sun2024auto}, which scaled up to millions of audio-caption pairs. However, this increase in scale came at the cost of descriptive quality. The data sources for these automated methods are often a limiting factor; for instance, WavCaps refines web-crawled text which is often coarse, while Auto-ACD and Sound-VECaps~\cite{yuan2025sound} rely on paired video data, restricting their applicability to audio-only contexts. 
Consequently, despite their size, these datasets often fail to resolve the core problem of descriptive granularity.

\begin{table*}[tb]
\caption{Performance of audio captioning models pre-trained on various datasets, evaluated on the MECAT-Caption benchmark. The final score is a weighted average of three main categories: systematic, Content-Specific, and Content-Unrelated. For all metrics, a higher score indicates better performance. Notably, a 'Pure' sample contains only one content type (e.g., only speech), while a 'Mixed' sample contains a combination of two or three types (i.e., speech, music, and sound events). $\dagger$ Combined refers to the combination of AudioSetCaps, Auto-ACD, WavCaps, and Sound-VECaps.}
\label{tab:mecat_caption_results}
\centering
\begin{tabular}{lcccccccccc}
\toprule
\multirow{4}{*}{Training Dataset} & \multicolumn{2}{c}{Systematic}   & \multicolumn{6}{c}{Content-Related}                                                                 & Content-Unrelated & \multirow{4}{*}{Score} \\ \cmidrule(lr){2-3} \cmidrule(lr){4-9} \cmidrule(lr){10-10}
                                 & Long           & Short          & \multicolumn{2}{c}{Speech}      & \multicolumn{2}{c}{Music}       & \multicolumn{2}{c}{Sound}       & Environment       &                        \\ \cmidrule(lr){4-5} \cmidrule(lr){6-7} \cmidrule(lr){8-9}
                                 &                &                & Pure           & Mixed          & Pure           & Mixed          & Pure           & Mixed          &                   &                        \\ \midrule
AudioSetCaps~\cite{bai2025audiosetcaps}                     & 52.4          & 52.0          & 30.2          & 31.4          & 44.3          & 30.9          & 52.4          & 21.6          & 15.4             & 37.4                  \\
Auto-ACD~\cite{sun2024auto}                          & 47.3          & 50.0          & 29.1          & 31.0          & 26.9          & 21.9          & 49.5          & 18.9          & 11.0             & 32.8                  \\
WavCaps~\cite{mei2024wavcaps}                          & 47.3          & 50.9          & 27.3          & 30.1          & 15.9          & 19.4          & 46.5          & 20.0          & 9.2     & 31.4                  \\
Sound-VeCaps~\cite{yuan2025sound}                    & 47.0          & 49.7          & 29.1          & 30.3          & 27.2          & 21.9          & 49.8          & 18.7          & 11.4             & 32.8                  \\ 
Combined\textsuperscript{$\dagger$} & 52.2 & 54.1 & 30.2 & 32.2 & 45.4 & 23.3 & 52.7 & 20.2 & 11.1 & 36.6 \\  \midrule
ACAVCaps                         & \textbf{76.6} & \textbf{75.7} & \textbf{64.2} & \textbf{64.9} & \textbf{60.5} & \textbf{41.1} & \textbf{59.5} & \textbf{28.0} & \textbf{34.8}             & \textbf{60.9}         \\ \bottomrule
\end{tabular}
\vspace{-3mm}
\end{table*}

The music domain shows a similar trajectory. Manually annotated datasets like MusicCaps~\cite{agostinelli2023music_caps} and SongDescriber~\cite{manco2023songdescriber} offer rich, detailed captions but are limited in scale. 
In response, automatically labeled datasets such as LPMusicCaps~\cite{doh2023lpmusiccaps} have emerged, leveraging LLMs to generate captions from existing musical databases. 
While larger, their descriptive granularity is often constrained by the richness of the source metadata, which may not contain the nuanced details of instrumentation, mood, and texture that a deep acoustic analysis could provide.

For speech, large-scale datasets have historically focused on tasks like automatic speech recognition (ASR), with their annotations capturing lexical content (transcripts) rather than descriptive captions of the acoustic scene. 
This leaves a significant gap, as the holistic description of a speech event -including tone, emotion, and environment — is crucial for general audio intelligence.

Thus, the current landscape of audio captioning datasets presents a trade-off between the high descriptive quality of small, manually-annotated datasets and the coarse granularity of larger, automatically-generated ones. 
This highlights a need for a resource that unifies large scale with fine-grained, acoustically-grounded descriptions across all major audio domains—sound events, music, and speech. 
A summary of our proposed dataset in comparison to existing ones can be seen in \Cref{tab:caption_comparison}.

\vspace{-3mm}
\section{ACAVCaps Data Construction}
\label{sec:methods}

The data construction pipeline is adapted from the methodology established in our prior work~\cite{niu2025mecat}, where a comprehensive description of the expert models and LLM prompts can be found. In this section, we provide a brief overview of this pipeline. As illustrated in \Cref{fig:labeling-main}, the multi-stage process is designed to capture a rich, multi-faceted understanding of each audio clip, beginning with analysis by a suite of specialized expert models and culminating in a final synthesis stage by a LLM.

\vspace{-1mm}
\subsection{Multi-Expert Annotation}
The initial analysis stage is designed to gather a comprehensive set of features from four key sources to inform the final caption generation. 
The primary source is a content-related analysis pipeline: a CED-Base model~\cite{dinkel2024ced} first classifies the audio to predict AudioSet~\cite{gemmeke2017audio} labels, which then routes the clip to specialized modules for speech (performing ASR and extracting speaker attributes), music (analyzing attributes like tempo and mood and separating vocals), or sound events (using the initial labels). 
This is supplemented by a content-unrelated analysis that universally characterizes acoustic properties such as signal intensity (RMS), recording quality, and reverberation. 
To provide further semantic context, we also generate a baseline description using a LALM and extract any original metadata, such as titles or tags, from the source file. 
Together, these structured analyses and raw metadata form the complete input for the final synthesis stage.





\subsection{LLM-CoT Reasoning}

The final stage leverages a LLM (Deepseek-R1~\cite{guo2025deepseek}) to synthesize the disparate outputs from the multi-expert analysis in conjunction with the original file metadata. 
Employing a Chain-of-Thought (CoT) prompting strategy, the LLM reasons over the collected evidence to resolve inconsistencies, infer relationships, and distill the most salient information. 
To ensure descriptive diversity, this process yields a final set of annotations where, for each identified acoustic scene or event, the LLM generates three semantically consistent yet stylistically varied captions. 
These are complemented by corresponding question-answer pairs, and all generated items are appended with a confidence score.

\begin{table*}[h]
\centering
\caption{Performance on downstream tasks. 
A model is pretrained on each respective training dataset. 
Then, for each task, we freeze the model and only optimize the adapter. 
For all speech tasks, lower is better, while for all other tasks higher is better. 
$\dagger$ Combined refers to the combination of AudioSetCaps, Auto-ACD, WavCaps, and Sound-VECaps.}
\label{tab:downstream_results}
\setlength{\tabcolsep}{5pt}
\begin{tabular}{lcccccccccc}
\toprule
\multirow{5}{*}{\begin{tabular}[c]{@{}l@{}}Training \\ Dataset\end{tabular}} & \multicolumn{6}{c}{Speech $\downarrow$} & \multicolumn{2}{c}{Sound $\uparrow$} & {Music $\uparrow$} & {Other $\uparrow$} \\ \cmidrule(lr){2-7} \cmidrule(lr){8-9} \cmidrule(lr){10-10} \cmidrule(lr){11-11}
& \multicolumn{3}{c}{AISHELL-2} & \multicolumn{2}{c}{LibriSpeech} & \begin{tabular}[c]{@{}c@{}}Common\\ Voice\end{tabular} & General & Vocal & Instrument & Emotion \\ \cmidrule(lr){2-4} \cmidrule(lr){5-6} \cmidrule(lr){7-7} \cmidrule(lr){8-9} \cmidrule(lr){10-10} \cmidrule(lr){11-11}
& Android & IOS & MIC & Clean & Other & French & VGGSound & VocalSound & NSynth & IEMOCAP \\ \midrule
AudioSetCaps & 82.7 & 77.8 & 81.7 & 51.6 & 70.2 & 84.7 & 22.4 & 91.4 & 67.0 & 17.6 \\
Auto-ACD & 89.1 & 78.2 & 88.6 & 54.6 & 76.5 & 85.7 & 22.5 & 90.2 & 46.1 & 24.1 \\
WavCaps & 83.2 & 74.2 & 77.9 & 54.3 & 74.0 & 85.2 & 21.2 & 91.5 & \textbf{69.1} & 19.9 \\
Sound-VECaps & 87.3 & 79.5 & 87.9 & 51.8 & 70.1 & 85.6 & 22.9 & 90.8 & 45.0 & 20.3 \\ 
Combined\textsuperscript{$\dagger$} & 84.2 & 76.4 & 82.3 & 41.5 & 59.4 & 83.0 & \textbf{34.6} & \textbf{92.6} & 44.0 & 19.8 \\ \midrule
ACAVCaps & \textbf{58.3} & \textbf{56.5} & \textbf{57.1} & \textbf{19.7} & \textbf{33.7} & \textbf{50.0} & 20.4 & 92.1 & 64.7 & \textbf{28.9} \\ \bottomrule
\end{tabular}
\vspace{-3mm}
\end{table*}

\vspace{-1mm}
\section{Experiments and Results}
\label{sec:exp_and_result}
To comprehensively evaluate our proposed dataset, ACAVCaps, we designed a series of experiments to assess the quality of the audio representations learned from it. 
Specifically, all models share a unified architecture consisting of a Dasheng-Base audio encoder~\cite{dinkel2024dasheng}, a lightweight MLP adapter, and a Qwen3-0.6B decoder~\cite{qwen3technicalreport}. 

\noindent\textbf{Implementation Details} All models were trained on eight GPUs. 
We used the AdamW8bit optimizer with a learning rate of $1\times 10^{-4}$ and a weight decay of 0.01 with a batch size of 16.
The training strategies, illustrated in \Cref{fig:labeling-main}, differed based on the evaluation task. 
For audio captioning, the audio encoder and MLP adapter were jointly trained while the LLM was fine-tuned with LoRA. 
Conversely, to assess downstream generalization, both the audio encoder and LLM were frozen, leaving only the MLP adapter trainable.

\subsection{Direct Performance on Audio Captioning}

To assess direct audio captioning performance, we conducted a comprehensive evaluation using the MECAT-Caption benchmark~\cite{niu2025mecat}. 
This benchmark provides a multi-faceted analysis of captioning quality across systematic, content-specific and content-unrelated categories, where performance is measured using the discriminative-enhanced audio text evaluation (DATE) score, a metric that rewards descriptive specificity in addition to semantic similarity. 

The results, presented in \Cref{tab:mecat_caption_results}, demonstrate the clear superiority of the model trained on ACAVCaps. 
It achieves an overall DATE score of 60.9, significantly outperforming models trained on other large-scale datasets. 
This validates that the superior detail fostered by ACAVCaps directly translates into an enhanced ability to generate fine-grained captions.

\vspace{-1mm}
\subsection{Analysis of Generalization Performance}
The unique token counts in \Cref{tab:caption_comparison} offer a static measure of each dataset's information richness. 
This section presents a more functional and dynamic evaluation, analyzing how that richness translates to the generalization capabilities of pre-trained models. 
Our evaluation is premised on the hypothesis that a dataset with greater informational breadth is key to learning more transferable representations. 
To validate this, we measure the downstream performance of models pre-trained on each dataset by fine-tuning them across four distinct and representative sub-domains of audio: speech content, sound events, music information, and paralinguistic attributes. 
For speech, we assess phonetic and linguistic understanding via multilingual ASR across Chinese (AISHELL-2~\cite{du2018aishell2}), English (LibriSpeech~\cite{panayotov2015librispeech}), and French (Common Voice~\cite{commonvoice}). 
For sound events, we test environmental awareness on a general classification benchmark (VGGSound~\cite{chen2020vggsound}) and the ability to distinguish non-speech human sounds (VocalSound~\cite{gong2022vocalsound}). 
In music, we gauge analytical proficiency through a fine-grained instrument recognition task (NSynth~\cite{nsynth2017}). 
Finally, to probe the understanding of paralinguistic attributes, we evaluate on a speech emotion recognition task (IEMOCAP~\cite{busso2008iemocap}). 
This comprehensive selection allows us to holistically evaluate the quality of the general-purpose representations that each dataset helps to cultivate.


The results in \cref{tab:downstream_results} substantiate our hypothesis. Notably, while the 'Combined' baseline comprises a larger sample size (6.0M vs. 4.7M), the ACAVCaps-trained model consistently exhibits superior performance across diverse downstream tasks. This performance gap is primarily attributed to informational density rather than sheer scale: despite having fewer audio-text pairs, ACAVCaps possesses a significantly higher unique token count (76.7K) compared to the Combined set (47.6K). Such lexical richness facilitates more precise audio-text alignment, fostering robust and generalizable representations. These findings illustrate that ACAVCaps achieves competitive or even superior results over larger aggregated datasets, reinforcing the premise that semantic complexity and data quality outweigh raw sample volume in audio-language pre-training."

\vspace{-3mm}
\section{Summary}
Progress in general audio understanding has been hampered by datasets limited in scale, scope, and descriptive detail. 
This paper addresses this bottleneck by introducing ACAVCaps, a large-scale audio captioning dataset designed to be comprehensive in content and diverse in its descriptive angles. 
Generated via a sophisticated pipeline, ACAVCaps provides a novel resource for training next-generation audio models.

Our experiments confirm that models trained on ACAVCaps exhibit superior performance. 
They not only excel at complex audio captioning tasks but also demonstrate strong generalization, successfully transferring to downstream speech, music, and sound event analysis tasks with significant improvements. 
This validates our core hypothesis: large-scale, comprehensive, and richly described datasets are crucial for developing robust and versatile audio representations.







\ninept
\bibliographystyle{IEEEbib}
\bibliography{strings}

\end{document}